# Between a ROC and a Hard Place: Using prevalence plots to understand the likely real world performance of biomarkers in the clinic.

Lendrem BC [1,3,4], Lendrem DW [1,2], Pratt AG [1,2,3], Naamane N [1,2], McMeekin P [5,6], Ng WF [1,2,3], Allen J [1,4], Power M [1,3,4] *, Isaacs JD [1,2,3]

[1] Institute of Cellular Medicine, Newcastle University, Newcastle upon Tyne, UK
[2] NIHR Newcastle Biomedical Research Centre, Newcastle University, Newcastle upon Tyne, UK
[3] Newcastle upon Tyne Hospitals NHS Trust, Newcastle upon Tyne, UK
[4] NIHR Newcastle In Vitro Diagnostics Co-operative, Newcastle University, Newcastle upon Tyne, UK
[5] Institute of Health & Society, Newcastle University, Newcastle upon Tyne, UK
[6] School of Health Sciences, Northumbria University, Newcastle upon Tyne, UK

**SUMMARY**

The Receiver Operating Characteristic (ROC) curve and the Area Under the Curve (AUC) of the ROC curve are widely used to compare the performance of diagnostic and prognostic assays. The ROC curve has the advantage that it is independent of disease prevalence. However, in this note we remind readers that the performance of an assay upon translation to the clinic is critically dependent upon that very same prevalence. Without an understanding of prevalence in the test population, even robust bioassays with excellent ROC characteristics may perform poorly in the clinic. Instead, simple plots of candidate assay performance as a function of prevalence rate give a more realistic understanding of the likely real-world performance and a greater understanding of the likely impact of variation in that prevalence on translational performance in the clinic.

**AUTHORSHIP**

All authors made substantial contributions to: (1) the conception and design of the study, or acquisition of data, or analysis and interpretation of data, (2) drafting the article or revising it critically for important intellectual content, (3) final approval of the version to be submitted.

**ROLE OF THE FUNDING SOURCE**

The authors were supported at least in part by the NIHR Newcastle Biomedical Research Centre and the NIHR In Vitro Diagnostics Cooperative at Newcastle University and the Newcastle upon Tyne Hospitals NHS Trust during the research and/or preparation of the article. The sponsors played no part in the study design; collection, analysis and interpretation of data; the writing of the report; or the decision to submit the article for publication.

**COMPETING INTERESTS**

The authors have no competing interests to declare.

The Receiver Operating Characteristic (ROC) curve is a widely used tool to evaluate diagnostic and prognostic biomarker performance[1,2,3]. The ROC curve compares the sensitivity and specificity of a candidate biomarker for a range of potential cut-off values for a biomarker assay. One of the perceived advantages of the ROC curve is that it is independent of the prevalence of the disease and captures the two key misclassification errors – false positive errors and false negative errors – as a function of biomarker cut-offs.

However, while the ROC curve is *independent* of the prevalence rate, the translational performance of a biomarker test in the clinic is *critically* dependent upon that very same prevalence rate[4,5]. For example, the "10-90-50 Rule" states that:

- for a disease with a prevalence of **10%**, and
- an assay with both sensitivity and specificity greater than **90%** (ROC AUC > 0.90),
- means that **50%** of patients testing positive are false alarms.

And if the prevalence of the disease is less than 10% then *most of our positive diagnostic tests will be false alarms* – see Figure 1.

While ROC AUCs and alternatives attempt to capture assay performance independently of prevalence, in this note we argue that there are advantages in looking at the robustness of a candidate assay to variation in those prevalence rates. Understanding how an assay performs across a range of values for the functional prevalence is critical in the clinic.

To begin with, there is often uncertainty surrounding the estimate of prevalence in the first place. Then, once the test is moved into the clinic, this is compounded by the fact that the prevalence rates vary depending upon how the patients are selected for testing. And, even following adoption of the test, the test may be used for groups of patients for whom the prevalence is rather less than that in the original test population, making the test virtually worthless. Translational performance is a function of both the 'true' disease prevalence *and the clinical selection process for testing*[4,5].

Rather than ignore prevalence, simple plots of candidate assay performance as a function of prevalence rate give a more realistic understanding of the likely real-world performance in the clinic, and a greater understanding of the likely impact of variation in that prevalence on translational performance in the clinic – see Figure 1B. Plotting the misclassification rates – *False Alarms* and *Missed Diagnoses* – as a function of possible prevalence rates allows us to focus on misclassification costs.

In Figure 2, we give a worked example showing prevalence plots for the promising mast cell activation test for IgE-mediated food allergy[6]. The sensitivity and specificity of this test are an impressive 97% and 92% respectively, with a ROC AUC of 0.99 (95% CI: 0.96, 1.00). While the number of patients with food allergies who are missed by the assay is reassuringly low, the number of patients without the disease testing positive is likely to be high, given an estimated prevalence in the UK of just 6%. This may, or may not, be acceptable. In real life, the relative costs associated with false alarms and missed diagnoses are likely to be very different and must be assessed prior to the test entering the clinic: a false alarm may simply mean a patient is subjected to further testing; a missed diagnosis may mean the patient dies.

Prevalence plots focus reviewers on misclassification rates, misclassification costs, and how the assay will translate to the clinic. Without thoughtful consideration of prevalence rates and the relative costs of misclassification errors, it is easy to 1) overstate the potential value of a candidate biomarker, 2) generate unrealistic expectations of that candidate, 3) incur unnecessary trial costs in evaluating that candidate, 4) incur opportunity costs in denying patients access to better diagnostic tests.

We provide an Excel workbook permitting readers to estimate Missed Cases, False Alarms and other key assay characteristics including prevalence plots for their assays for any given sensitivity and specificity.

A

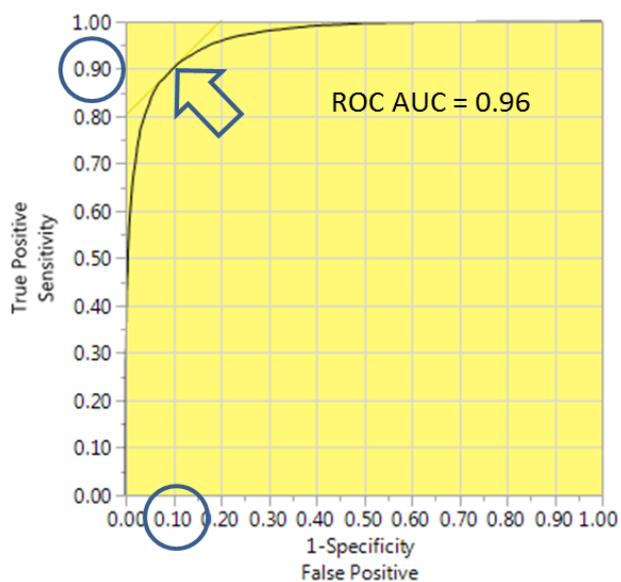

B

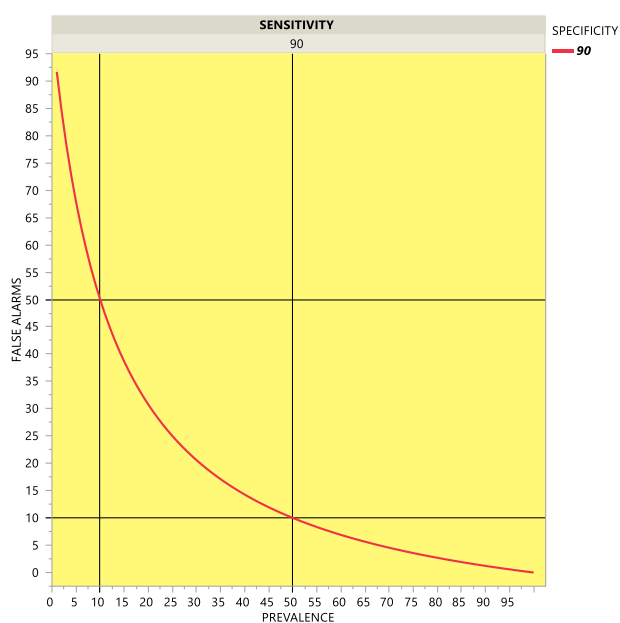

**Figure 1: The 10-90-50 Rule.** While the ROC AUC for our candidate biomarker looks promising (**A**) giving an assay with both sensitivity and specificity of 90%, the performance of the assay in the clinic depends critically on the prevalence of the disease (**B**). The false positive and false negative rates are both 10%, but if the prevalence of the disease in the test population is 10% then 50% of all positive tests will be false alarms. The false alarm rate depends critically upon the prevalence in the test population. Plotting test performance as a function of prevalence gives a more realistic understanding of likely performance in the clinic. See text for details.

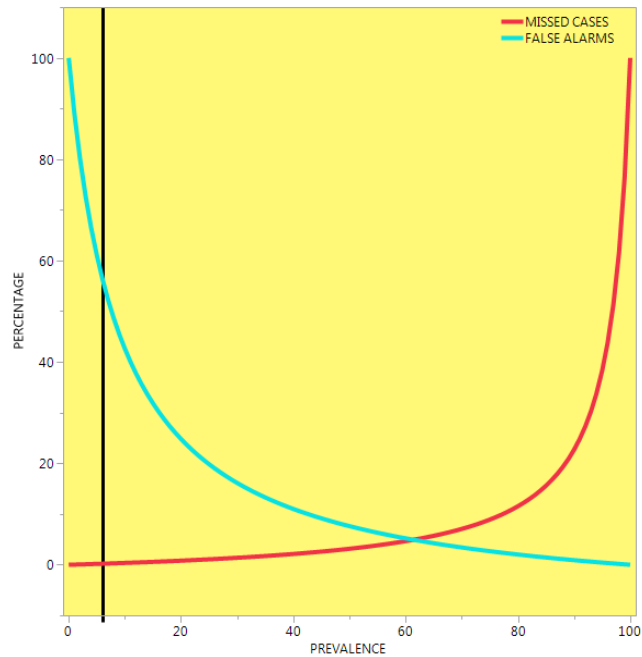

**Figure 2: Prevalence plots for the mast cell activation assay.** While the assay looks promising - with a sensitivity of 97%, specificity of 92% and a ROC AUC of 0.99 - translation to the clinic depends critically upon the prevalence in the test population. As the prevalence increases, the percentage of missed cases increase and the false alarms decrease. If the prevalence rate is zero then any positive test results are false positives and the false alarm rate is 100%. If the prevalence rate is 100% then any negative tests are false negatives and the missed case rate is 100%. *The vertical line shows the estimated prevalence of IgE-mediated food allergy at 6%.* At this rate, while 56% of all positive tests will be false alarms, just 17 tests will be needed to identify each new case of IgE-mediated food allergy – see **Supplementary Excel Workbook**.